\documentclass[aip,graphicx]{revtex4-1}
\usepackage{graphicx}
\draft 

\begin{document}


\title{Beyond Impedance Additivity: A Systematic Nonlinear Perspective on Memristor Associations} 



\author{Luiz A. Meneghetti Jr. }
 \email{luizmeneghetti7@gmail.com}
\affiliation{Department of Physics, Federal University of São Carlos, 13565-905 São Carlos, SP, Brazil}

\author{Leonardo K. Castelano}
\affiliation{Department of Physics, Federal University of São Carlos, 13565-905 São Carlos, SP, Brazil}

\author{Antônio Sérgio dos Santos}
\affiliation{Department of Physics, Federal University of São Carlos, 13565-905 São Carlos, SP, Brazil}

\author{Soumen Pradhan}
\affiliation{Julius-Maximilians-Universität Würzburg, Physikalisches Institut and Würzburg-Dresden Cluster of Excellence ctd.qmat, Lehrstuhl für Technische Physik, Am Hubland, 97074 Würzburg, Germany}

\author{Fabian Hartmann}
\affiliation{Julius-Maximilians-Universität Würzburg, Physikalisches Institut and Würzburg-Dresden Cluster of Excellence ctd.qmat, Lehrstuhl für Technische Physik, Am Hubland, 97074 Würzburg, Germany}

\author{Ovidiu Lipan}
\affiliation{Department of Physics, University of Richmond, 28 Westhampton Way, Richmond, Virginia 23173, USA}

\author{Sven Höfling}
\affiliation{Julius-Maximilians-Universität Würzburg, Physikalisches Institut and Würzburg-Dresden Cluster of Excellence ctd.qmat, Lehrstuhl für Technische Physik, Am Hubland, 97074 Würzburg, Germany}
 
\author{Victor Lopez-Richard}
 \email{vlopez@df.ufscar.br}
\affiliation{Department of Physics, Federal University of São Carlos, 13565-905 São Carlos, SP, Brazil}



\begin{abstract}
We investigate the validity of the superposition principle and impedance additivity in AC circuits containing a memristive device connected in series with a resistor, capacitor, or inductor. While the series association of impedances is a cornerstone of linear circuit theory, its applicability to memory-bearing nonlinear systems remains largely unexplored. Using a state-dependent memristive model, we numerically analyze the stationary current response under sinusoidal excitation and characterize the resulting harmonic spectra, Bode diagrams, and Nyquist plots. To assess whether the fundamental response can still be interpreted through an equivalent-circuit framework, we introduce the concept of an apparent memristor, whose effective parameters are extracted directly from the composite impedance. We show that, although the fundamental harmonic can be accurately reproduced by an apparent equivalent circuit over selected parameter ranges, the effective parameters differ substantially from those of the isolated memristor, revealing a renormalization induced by the coupling to the passive element. More importantly, we identify parameter regimes in which the apparent-circuit description breaks down altogether, particularly for capacitor- and inductor-coupled systems, demonstrating that the composite impedance cannot generally be expressed as the sum of independent impedances. These results establish boundaries for the use of equivalent-circuit models in memory-enabled electronic systems and provide practical guidelines for the interpretation of impedance spectroscopy in nonlinear devices exhibiting memory. 
\end{abstract}


\maketitle 

\section{\label{sec:level1} Introduction }
The interaction between memristive devices and passive circuit elements has attracted considerable attention because external components not only modify the electrical response of the circuit, but may also influence the internal state evolution responsible for memory. In particular, the simplest hybrid configuration, consisting of a memristor connected in series with a resistor, has been extensively investigated. An analog implementation employing a memristor emulator and a variable resistor was presented in Ref.~\cite{2010-Pershian}, demonstrating how the external resistance modifies the shape of the pinched hysteresis loop and the overall device response. A comprehensive theoretical treatment of the same configuration was later developed in Ref.~\cite{2023-Kitaev}, where the coupled nonlinear dynamics of the circuit were analyzed through differential equations and compared with experimental observations. These studies clearly demonstrate that the series resistor is not a passive spectator but actively alters the memristor dynamics. Similar conclusions have also been reported in Refs.~\cite{2008-Makoto,2021-Maldonado,2024-Bisquert}, which show that increasing the external resistance progressively suppresses the characteristic hysteresis of the memristive response.

The influence of reactive elements has also been widely explored. For memristor--capacitor circuits, the pioneering work of Ref.~\cite{2009-Joglekar} investigated the dynamics of an ideal memristor connected in series with a capacitor, while Ref.~\cite{2015-Mutlu} exploited the same configuration to determine the operating frequencies of memristive relaxation oscillators. Fractional-order extensions were subsequently considered in Ref.~\cite{2020-WANG}, whereas Ref.~\cite{2022-Slipko} established a broader theoretical framework for describing mixed circuits composed of stochastic memristive and reactive elements. More recently, Ref.~\cite{2022-Fang} employed the memristor--capacitor configuration as a building block for modeling neuronal dynamics. The effect of capacitive coupling on the hysteretic response of memory devices has also been discussed in Refs.~\cite{2020-WANG,2023-Bisquert,2023-Bisquert-PRA,2024-Bisquert}.

The complementary memristor--inductor configuration has likewise received considerable attention. The classical work of Ref.~\cite{2009-Joglekar} analyzed the temporal response of both memristor--capacitor and memristor--inductor circuits, highlighting the role of resistive memory in modifying the circuit dynamics. A more detailed dynamical treatment was later presented in Ref.~\cite{2023-Kitaev}. Furthermore, while Ref.~\cite{2009-Joglekar} investigated hysteresis arising from the interaction between a memristor and an external inductor, Ref.~\cite{2023-Bisquert} showed that inductive signatures may also emerge intrinsically from the dynamics of memristive devices themselves.

Despite these extensive investigations, a fundamental question has received comparatively little attention. In conventional linear AC circuit theory, the response of elements connected in series is described through the principle of superposition, according to which the total impedance is simply the sum of the individual impedances. This principle underlies virtually all equivalent-circuit analyses and forms the basis of impedance spectroscopy. However, when one of the circuit elements possesses an internal state that evolves according to the applied excitation, the dynamics of the passive element and the memory variable become intrinsically coupled. Under these conditions, it is no longer evident whether the impedance of the composite circuit can still be interpreted as the sum of independent contributions or whether the coupling fundamentally renormalizes the apparent electrical properties of the memristive element.

The objective of the present work is precisely to address this question. Rather than focusing exclusively on how passive elements modify hysteresis, we investigate the validity of impedance additivity in hybrid circuits composed of a memristive device connected in series with a resistor, capacitor, or inductor. By comparing the exact nonlinear response of the composite system with an apparent equivalent-circuit description, we identify the parameter regimes in which conventional equivalent circuits remain meaningful and those in which the breakdown of the superposition principle becomes unavoidable. In doing so, we establish a systematic framework for quantifying how memory effects renormalize the apparent impedance of nonlinear electrical systems, providing practical guidelines for the interpretation of impedance spectroscopy and equivalent-circuit modeling of memory-enabled electronic devices.

\section{\label{sec:methods} Method and Results}

Due to the assumptions of linearity and time invariance that underlie conventional AC circuit analysis, the overall response of a circuit can be determined through the principle of superposition. Within this framework, Kirchhoff’s laws lead directly to impedance additivity, such that the total voltage across a series network is given by the sum of the voltage drops across the individual elements,
$Z_n$, $V=(Z_1+Z_2 +Z_3+...)I$~\cite{2018-Dorf},where $Z_n$ denotes the impedance of the $n$-th circuit element and $I$ is the current flowing through the system~\cite{2018-Dorf}. This relation forms the basis of equivalent-circuit representations and frequency-domain analysis in linear electrical networks.

However, in systems exhibiting memory, such as memristors and many photovoltaic devices, the validity of this relation becomes questionable. In these systems, the electrical response depends not only on the instantaneous voltage or current but also on the previous evolution of the system, introducing history-dependent dynamics that are absent in conventional linear elements~\cite{1971-Chua,2008-Strukov,2013-Ventura,2021-GAO,2020-Corinto,2025-Nogueira}. As a consequence, the voltage–current relationship becomes nonlinear and state-dependent, challenging the assumptions required for impedance additivity and superposition. In this work, we investigate circuits composed of memristive devices, more generally, systems exhibiting conductive memory responses, connected in series with passive linear elements. Our objective is to assess the extent to which equivalent-impedance descriptions remain valid beyond the linear regime and to identify the conditions under which they fail. More broadly, we seek to establish a systematic framework for quantifying deviations from impedance additivity, thereby providing practical guidelines for incorporating memory effects into the analysis and interpretation of nonlinear electrical circuits.

Our memristive model is based on the assumption that the device conductance is controlled by the generation of a nonequilibrium population of charge carriers, denoted by $\delta N(t)$, which may originate from a variety of microscopic mechanisms, including charge trapping, ionic migration, defect redistribution, or polarization-induced carrier modulation~\cite{Paiva2022,2015-Messerchemit,2024-LopezRichard-JAP}. The temporal evolution of this carrier imbalance is described by
\begin{equation}
    \label{eq:delta_n}
    \frac{d\delta N(t)}{dt} = - \frac{\delta N(t)}{\tau} + g(V),
\end{equation}
 where $\tau$ is the characteristic relaxation time (used here as the unit of time, $\tau=1$), $g(V)$ is a voltage-dependent generation function, and $\delta N(0)$ specifies the initial nonequilibrium state of the system~\cite{2011-Pershin,2009-Biolek,2013-Biolek,2024-LopezRichard-PRB,2024-LopezRichard-JAP}. This equation captures the competition between carrier generation and relaxation, providing a minimal framework for describing memory effects in conductive systems.

When the memristive device is connected in series with passive circuit elements, the voltage drop across the device, $V_M(t)$, directly determines the generation rate entering Eq.~\ref{eq:delta_n}. For analytical simplicity, the generation function is expanded up to second order in $V_M(t)$
\begin{equation}
\label{eq:generatingfunc}
g(V_M) = \sigma_o V_M(t) + \sigma_e V_M^2(t),
\end{equation}
where $\sigma_o$ and $\sigma_e$ characterize the odd- and even-voltage contributions, respectively. These coefficients can be associated with distinct microscopic processes and symmetry properties of the underlying transport mechanisms, as discussed in Refs.~\citenum{2009-LopezRichard,Paiva2022,2024-LopezRichard-JAP}.

The electrical current through the memristive device is assumed to depend linearly on the nonequilibrium carrier population, yielding
\begin{equation}
\label{eq:currentvm}
I(t) = [G_0 + \gamma \delta N(t)]\, V_M(t),
\end{equation}
where $G_0$ is the equilibrium conductance and $\gamma$ quantifies the sensitivity of the conductance to the carrier imbalance, thereby setting the strength of the memory response~\cite{2022-Silva}.

For a memristive device connected in series with a resistor of resistance $R$, current continuity immediately gives
 \begin{equation}
 V_M(t)=\frac{V(t)}{1+R[G_0 +\gamma \delta N(t)]}.\label{eq:VMR}
 \end{equation}
where $V(t)$ is the externally applied voltage. In this configuration, the resistor acts as a voltage divider whose ratio evolves dynamically according to the memristive state.

For a memristive device connected in series with a capacitor, the voltage across the device is
\begin{equation}
\label{eq:VMC}
V_M(t) = V(t) - \frac{q(t)}{C},
\end{equation}
where $C$ is the capacitance and $q(t)$ is the charge accumulated on the capacitor. In this case, the memory dynamics of the device becomes coupled to the charge-storage dynamics of the capacitor.

Finally, for a memristive device connected in series with an inductor, the voltage drop across the device satisfies
\begin{equation}
\label{eq:VML}
V_M(t) = V(t) - L \frac{dI(t)}{dt},
\end{equation}
where $L$ is the inductance. The inductor introduces an additional dynamical timescale associated with the inertia of the current, leading to a direct interplay between inductive and memristive effects.

Combining Eqs.~\ref{eq:delta_n}-\ref{eq:VML} yields the governing equations for the three composite systems. For the memristor–resistor configuration, the dynamics is described by a single nonlinear ordinary differential equation,
 \begin{equation}
    \label{eq:ODEMR}
    \frac{d\delta N(t)}{dt} = - \frac{\delta N(t)}{\tau} + g\left(\frac{V}{1+R[G_0+\gamma\delta N]}\right).
\end{equation}
which highlights how the resistor modifies the effective driving voltage experienced by the memristive element.

For the memristor–capacitor configuration, the system is governed by two coupled ordinary differential equations,
\begin{eqnarray}
    \frac{d\delta N(t)}{dt} &=& - \frac{\delta N(t)}{\tau} + g\left(V-\frac{q}{C}\right),\\
    \frac{dq(t)}{dt} &=& (G_0+\gamma\delta N)\left(V-\frac{q}{C}\right) ,
    \label{eq:ODEMC}
\end{eqnarray}
revealing the mutual coupling between the carrier-imbalance dynamics and the capacitor charge.

Similarly, for the memristor–inductor configuration, one obtains
\begin{eqnarray}
    \frac{d\delta N(t)}{dt} &=& - \frac{\delta N(t)}{\tau} + g\left(\frac{I}{G_0+\gamma\delta N}\right),\\
    \frac{dI(t)}{dt} &=& \frac{1}{L}\left[V-\frac{I}{(G_0+\gamma\delta N)}\right],
    \label{eq:ODEML}
\end{eqnarray}
where the current dynamics and the internal memristive state evolve self-consistently. Together, these equations provide a unified framework for investigating how passive circuit elements modify the nonlinear and history-dependent response of memristive systems, and for assessing the conditions under which conventional impedance-addition concepts cease to be valid.

The systems described by Eqs.~\ref{eq:delta_n}--\ref{eq:ODEML} were solved numerically under a sinusoidal excitation $V(t)=V_0\cos(\omega t)$. To ensure that the reported results correspond to the stationary regime, the temporal evolution was followed over successive driving cycles until all transient contributions vanished. Convergence was assessed by comparing the current response (I(t)) in consecutive periods. Specifically, the stationary state was considered reached when the integrated difference between the current profiles of two successive cycles, normalized by the current amplitude, became smaller than $10^{-6}$. Unless otherwise stated, all simulations were performed using $G_0=1$, $\gamma=1$, $\sigma_o=0$, and $\sigma_e=3.7$. This parameter set produces the characteristic pinched hysteresis loop expected for an isolated memristive device, providing a convenient reference for evaluating the influence of the passive circuit elements.

\begin{figure}[t]
    \centering
    \includegraphics[width=0.7\textwidth]{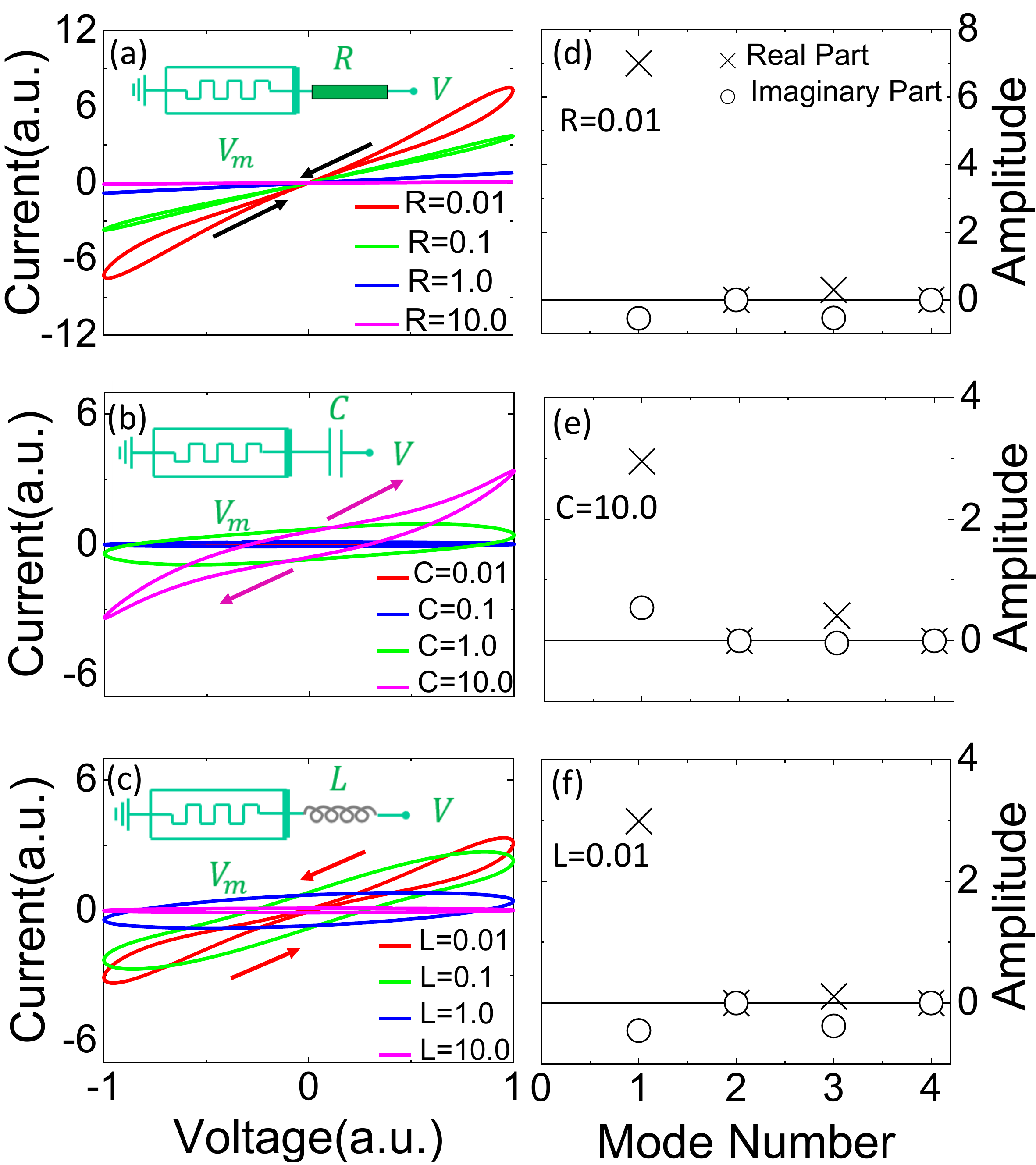}
    \caption{\label{fig:histerese}
     Characterization of the memristor response under series association with external passive parameters. Subfigures (a), (b), and (c) illustrate the hysteresis curves obtained when the memristor is coupled to varying external resistance ($R$), capacitance ($C$), and inductance ($L$), respectively. In these cases, the values of the external parameters are systematically varied by multiplying them by successive factors of 10 (as indicated in the legend), serving as control variables to analyze their influence on the nonlinear current–voltage characteristics. Subfigures (d), (e), and (f) present the corresponding modal distributions for fixed parameter values of $R=0.01$, $C=10.0$, and $L=0.01$, highlighting the role of these external parameters in shaping the dynamic and modal behavior of the system.
    }
\end{figure}

Figure~\ref{fig:histerese}(a) displays the stationary $I-V$ characteristics of a memristive device connected in series with a resistor for representative values of $R=0.01$, $0.1$, $1.0$, and 10 (in units of $G_0^{-1}$). For small resistances ($R<0.1$), the voltage drop across the resistor remains negligible, and the response is dominated by the intrinsic nonlinear dynamics of the memristive element, preserving the characteristic pinched hysteresis loop. As $R$ increases ($0.1\leq R<10$), a progressively larger fraction of the applied voltage is dissipated in the resistor, reducing the effective voltage acting on the memristor. Consequently, the hysteresis area shrinks and the nonlinear features become less pronounced. In the strongly resistive regime ($R\geq10$), the memristive contribution becomes largely suppressed, and the $I-V$ characteristic approaches a straight line, signaling the recovery of an almost purely resistive response.

Figure~\ref{fig:histerese}(b) presents the corresponding results for a memristive device connected in series with a capacitor. For low capacitances ($C\leq1$, in units of $\tau G_0$), the capacitive reactance dominates the circuit response, producing nearly elliptical $I-V$ curves characteristic of a conventional capacitor. In this regime, the phase shift introduced by charge accumulation overwhelms the intrinsic memory response of the device. As the capacitance increases, the capacitive impedance decreases and the influence of the memristor becomes progressively more evident. For intermediate values, such as $C=10$, both mechanisms contribute on comparable footing, giving rise to distorted loops in which the pinched memristive hysteresis coexists with the elliptical signature of a reactive element. This crossover regime provides a clear manifestation of the competition between conductive memory and capacitive charge-storage dynamics.

The behavior of a memristive device connected in series with an inductor is shown in Fig.~\ref{fig:histerese}(c). For small inductances ($L \leq 0.01$, in units of $\tau/G_0$), the inductive reactance is negligible compared to the effective memristive resistance, and the current closely follows the voltage excitation. As a result, the characteristic memristive hysteresis is largely preserved. Increasing the inductance ($0.1\leq L\leq 1.0$) introduces a substantial phase lag between current and voltage, progressively deforming the pinched loop into a broader, oval-shaped trajectory. In this intermediate regime, the system dynamics is determined by the interplay between the memory-dependent conductance of the memristor and the current inertia imposed by the inductor. For large inductances ($10\leq L\leq 100$), the inductive reactance becomes the dominant contribution, severely limiting the rate at which the current can vary. The resulting $I-V$ characteristics collapse into nearly horizontal trajectories, reflecting a response governed primarily by inductive constraints rather than by the intrinsic memory properties of the device.

Across all three configurations, a well-defined hysteresis loop emerges whenever the memristive contribution dominates the composite response. This behavior demonstrates that the memory-dependent dynamics remain robust even in the presence of passive circuit elements. More importantly, the results reveal that the interaction between memory effects and passive components cannot, in general, be interpreted as a simple superposition of independent responses. Depending on the relative values of $R$, $C$, and $L$, the memristive dynamics can substantially modify the overall circuit behavior, including regimes in which the passive element would traditionally be expected to dictate the response.

\begin{figure}[t]
    \centering
    \includegraphics[width=0.9\textwidth]{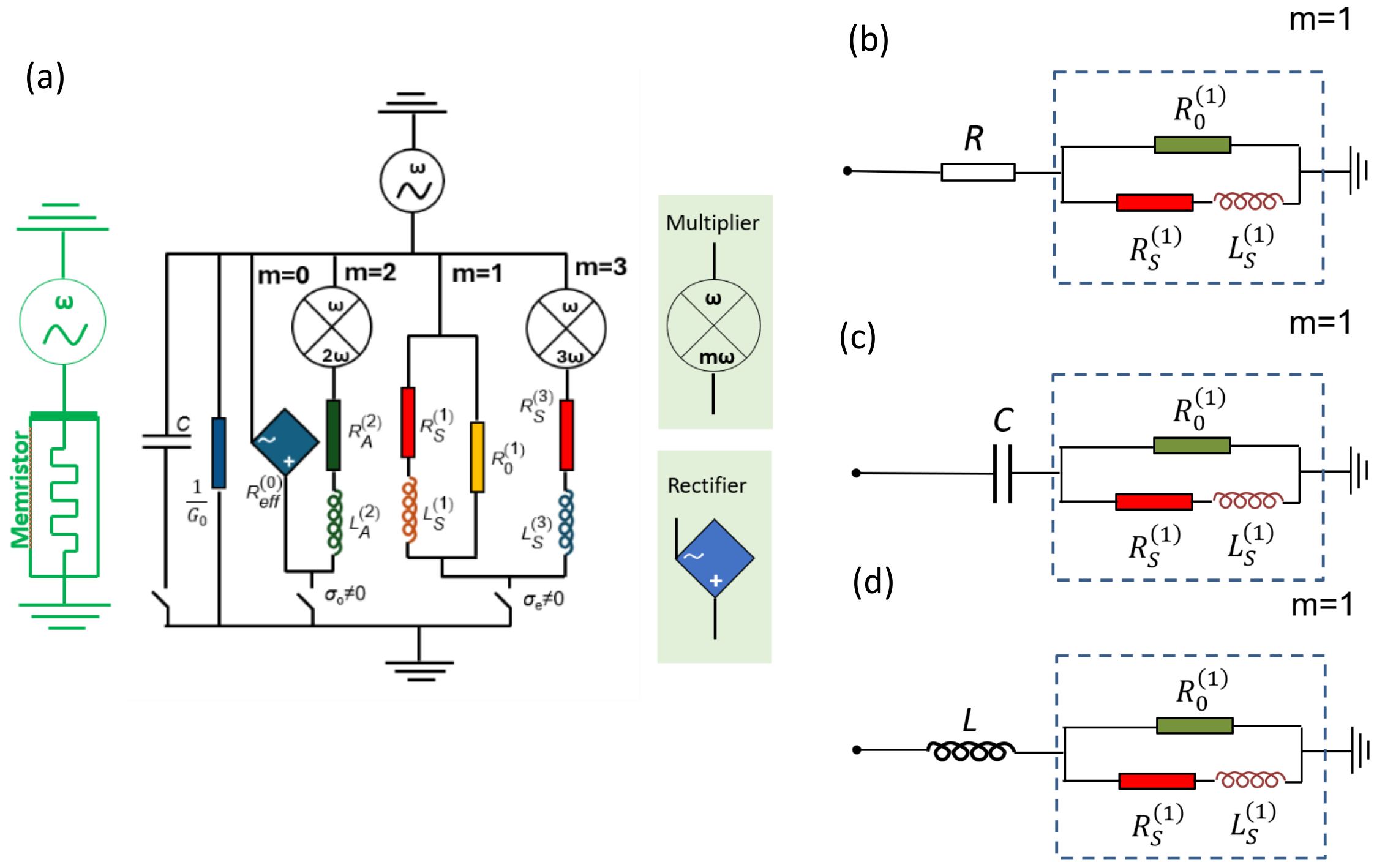}
    \caption{\label{fig:diagram}
    Representation of the equivalent circuit of the memristor obtained from the Fourier series expansion. Subfigure (a) shows the complete equivalent circuit considering the first four harmonic modes ($m = 0–3$). Subfigures (b)–(d) present the equivalent circuit corresponding only to the fundamental mode ($m = 1$), coupled to different external passive elements: (b) resistor, (c) capacitor and (d) inductor.
    }
\end{figure}

To gain further insight into the nonlinear dynamics of the composite systems, the stationary current is expanded as a Fourier series~\cite{2012-Joglekar,2024-LopezRichard-JAP}
\begin{equation}
    \label{eq:harmonic I}
    I(t)= \sum_{m=1} [G_{(m)} \cos(m\omega t) - B_{(m)} sin(m\omega t)],
\end{equation}
where $G_{(m)}$ and $B_{(m)}$ denote the conductance and susceptance associated with the $m$-th harmonic mode, respectively~\cite{2012-Joglekar,2024-LopezRichard-JAP,2023-LopezRichard,2024-LopezRichard-PRB}. This decomposition provides direct access to the spectral signatures of nonlinear and memory-driven processes. In particular, the appearance of higher-order harmonics ($m>1$) constitutes a direct manifestation of nonlinearity, since purely linear systems driven by a sinusoidal excitation respond exclusively at the driving frequency~\cite{2024-LopezRichard-JAP,2008-Makoto}.

The relative contributions of the different harmonic modes for the memristor connected in series with a resistor, capacitor, and inductor are shown in Figure~\ref{fig:histerese}(d), (e), and (f). Figure~\ref{fig:histerese}(d) presents the harmonic spectrum for the memristor–resistor configuration with $R=0.01$. In this regime, the real part of the first harmonic ($m=1$) is dominant, reflecting the fact that the response remains largely governed by the fundamental excitation. While the third harmonic ($m=3$) contributes only weakly to the overall response, its nonzero amplitude reveals the existence of nonlinear processes that cannot be captured within a linear impedance framework. 

A different scenario emerges for the memristor–capacitor configuration, shown in Fig.~\ref{fig:histerese}(e) for $C=10$. Although the first harmonic remains the dominant contribution, the corresponding susceptance becomes positive, revealing the phase advance of the current relative to the applied voltage that is characteristic of capacitive systems. At the same time, the higher-order harmonics exhibit noticeably larger amplitudes than in the resistive case, indicating a stronger departure from purely sinusoidal behavior. This enhanced harmonic content reflects the intricate coupling between the charge-storage dynamics of the capacitor and the history-dependent conductance of the memristive element, giving rise to a rich nonlinear response.

Figure~\ref{fig:histerese}(f), presents the harmonic spectrum for the memristor–inductor configuration with $L=0.01$. The presence of the inductor introduces a negative susceptance component, reflecting the phase lag of the current relative to the applied voltage. As a result, the current dynamics become coupled to the evolution of the memristor's internal state. This interplay between inductive inertia and memory-dependent conductance enhances the nonlinear character of the response, leading to the contribution from higher-order harmonics.

To investigate how the passive element modifies the frequency response of the composite system, we analyze the impedance through its Bode representation. Since the current is expressed as the Fourier expansion of Eq.~\ref{eq:harmonic I}, each harmonic mode can be associated with a complex impedance,
\begin{equation}
    \label{eq:impedance}
    Z_{(m)}=\frac{1}{G_{(m)} +i B_{(m)}},
\end{equation}
where $Z_{(m)}$ denotes the impedance associated with the $m$-th harmonic. Unlike linear systems, however, each harmonic represents an independent contribution to the nonlinear response, and therefore the concept of a single circuit impedance is generally no longer applicable.

Our previous work demonstrated that an isolated memristor driven by a sinusoidal voltage, $V(t)=V_0\cos{\omega t}$, can be represented in terms of its harmonic modes~\cite{2024-LopezRichard-JAP}. In particular, the fundamental mode ($m=1$) is exactly equivalent to the response of the circuit shown in Fig.~\ref{fig:diagram}(a), consisting of a resistor connected in parallel with a resistor-inductor series branch. This equivalent circuit provides an effective description of the first harmonic, even though the complete memristive response remains nonlinear.

A natural question then arises: can an analogous equivalent circuit still be defined when the memristor is connected in series with a passive element? If impedance additivity remained valid, the answer would be straightforward. The fundamental-mode impedance of the composite circuit would simply be 
\begin{equation}
Z^{M}_{(1)}+Z_{p},
\end{equation}
where $Z^{M}_{(1)}$ is the fundamental impedance of the isolated memristor and $Z_{p}$ is the impedance of the resistor, capacitor, or inductor. The corresponding apparent circuits are illustrated in Fig.~\ref{fig:diagram} panels (b-d).

However, the nonlinear coupling between the passive element and the internal state of the memristor invalidates this simple addition. Consequently, the impedance extracted from the composite circuit generally cannot be interpreted as the sum of two independent impedances. Instead, we ask a more practical question: to what extent can the first harmonic still be described by the same circuit topology if its parameters are allowed to change?

To answer this question, we introduce an effective memristor impedance 
\begin{equation}
Z_{\mathrm{eff}}=Z_{(1)}-Z_p, 
\end{equation}
where the subtraction of the passive-element impedance serves only as a formal definition. Importantly, $Z_{\mathrm{eff}}$ should not be interpreted as the impedance of the isolated memristor. Rather, it represents the apparent impedance that the memristive element would need to possess in order for the first harmonic of the composite system to be reproduced by the same equivalent-circuit topology. Any deviation of the extracted parameters from those of the isolated device therefore quantifies the failure of the superposition principle and the renormalization produced by the coupling to the passive element.

Assuming that the topology of the equivalent circuit is preserved, its effective impedance can be written as
\begin{equation}
    \frac{1}{Z_{\mathrm{eff}}}=\left(G_{\mathrm{eff}}
    + \frac{S_{\mathrm{eff}}}{2}\right)
    + \frac{S_{\mathrm{eff}}}{4} \left(
        \frac{1}{1 + (2\omega \tau_{\mathrm{eff}})^2}
    \right) - i \frac{S_{\mathrm{eff}}}{4}
    \left(
        \frac{2\omega \tau_{\mathrm{eff}}}{1 + (2\omega \tau_{\mathrm{eff}})^2}
    \right).
\end{equation}
The parameters $G_{\mathrm{eff}}$, $S_{\mathrm{eff}}$, and $\tau_{\mathrm{eff}}$ are therefore effective (or apparent) parameters, rather than intrinsic material quantities. They characterize the response of the composite circuit as perceived through its fundamental harmonic and provide a quantitative measure of how the coupling to the passive element renormalizes the apparent properties of the memristive device.

The real and imaginary parts of the effective impedance of the equivalent circuit representation are
\begin{equation}
\label{eq:imp_real}
    Z_{\mathrm{eff}}^{(\mathrm{Re})}= \frac{
        4\left[1 + (2\omega\tau_{\mathrm{eff}})^2\right]
        \left[4(G_{\mathrm{eff}} + \frac{S_{\mathrm{eff}}}{2})\left(1 + (2\omega\tau_{\mathrm{eff}})^2\right)\right]
    } {
        4\left(G_{\mathrm{eff}} + \frac{S_{\mathrm{eff}}}{2}\right)
        \left(1 + (2\omega \tau_{\mathrm{eff}})^2\right)
        + (2\omega \tau_{\mathrm{eff}})^2 S_{\mathrm{eff}}^{\,2}
}
\end{equation}
and
\begin{equation}
\label{eq:imp_im}
Z_{\mathrm{eff}}^{(\mathrm{Im})}
    =
    \frac{
        8\left[1 + (2\omega \tau_{\mathrm{eff}})^2\right]\,(\omega\tau_{\mathrm{eff}} S_{\mathrm{eff}})
    }
    {%
        4\left(G_{\mathrm{eff}} + \frac{S_{\mathrm{eff}}}{2}\right)
        \left(1 + (2\omega \tau_{\mathrm{eff}})^2\right)
        + (2\omega \tau_{\mathrm{eff}})^2 S_{\mathrm{eff}}^{\,2}
    }
\end{equation}

These expressions enable the effective parameters to be extracted directly from the frequency response. In particular, $S_{\mathrm{eff}}$ and $G_{\mathrm{eff}}$ are determined from the low- and high-frequency limits of the real part of the impedance, in Eq.~\ref{eq:imp_real}
\begin{equation}
    S_{\mathrm{eff}}
    =
    4\left(
        \frac{1}{\lim_{\omega \to 0} Z_{\mathrm{eff}}^{(\mathrm{Re})}}
        -
        \frac{1}{\lim_{\omega \to \infty} Z_{\mathrm{eff}}^{(\mathrm{Re})}}
    \right),\label{eq:seff}
\end{equation}
and
\begin{equation}
    G_{\mathrm{eff}}
    =
    \frac{3}{\lim_{\omega \to \infty} Z_{\mathrm{eff}}^{(\mathrm{Re})}}
    -
    \frac{2}{\lim_{\omega \to 0} Z_{\mathrm{eff}}^{(\mathrm{Re})}}.\label{eq:geff}
\end{equation}
The effective relaxation time is obtained by combining the real and imaginary components of the impedance. Denoting by $\omega_{\mathrm{max}}$ the frequency at which $Z_{\mathrm{eff}}^{(\mathrm{Im})}$ reaches its maximum, one finds
\begin{equation}
    \tau_{\mathrm{eff}}
    =
    \frac{1}{\omega_{\mathrm{max}}}
    \frac{4G_{\mathrm{eff}} + 3S_{\mathrm{eff}}}{2G_{\mathrm{eff}} + S_{\mathrm{eff}}}.\label{eq:teff}
\end{equation}

Finally, the extracted effective parameters can be mapped onto the apparent equivalent circuit shown in Fig.~\ref{fig:diagram}, yielding 
\begin{equation}
    R^{\mathrm{eff}}_0 = \frac{1}{G_{\mathrm{eff}} + \frac{S_{\mathrm{eff}}}{2}},\label{eq:reff0}
\end{equation}
\begin{equation}
    R^{\mathrm{eff}}_S = \frac{4}{S_{\mathrm{eff}}},\label{eq:reffs}
\end{equation}
and
\begin{equation}
    L^{\mathrm{eff}}_S = 8\,\frac{\tau_{\mathrm{eff}}}{S_{\mathrm{eff}}}, \label{eq:Leffs}
\end{equation}
where $R_0^{\mathrm{eff}}$ corresponds to the resistor connected in parallel with the series branch formed by $R_S^{\mathrm{eff}}$ and $L_S^{\mathrm{eff}}$. These quantities should be interpreted as apparent circuit parameters that reproduce the fundamental-mode response of the composite system. Their dependence on the values of $R$, $C$, or $L$ provides a direct measure of the extent to which the coupling to passive elements modifies the apparent memristive behavior and, consequently, of the validity of equivalent-circuit descriptions based on impedance additivity.

\begin{figure}[t]
	\includegraphics[width=1.0\textwidth]{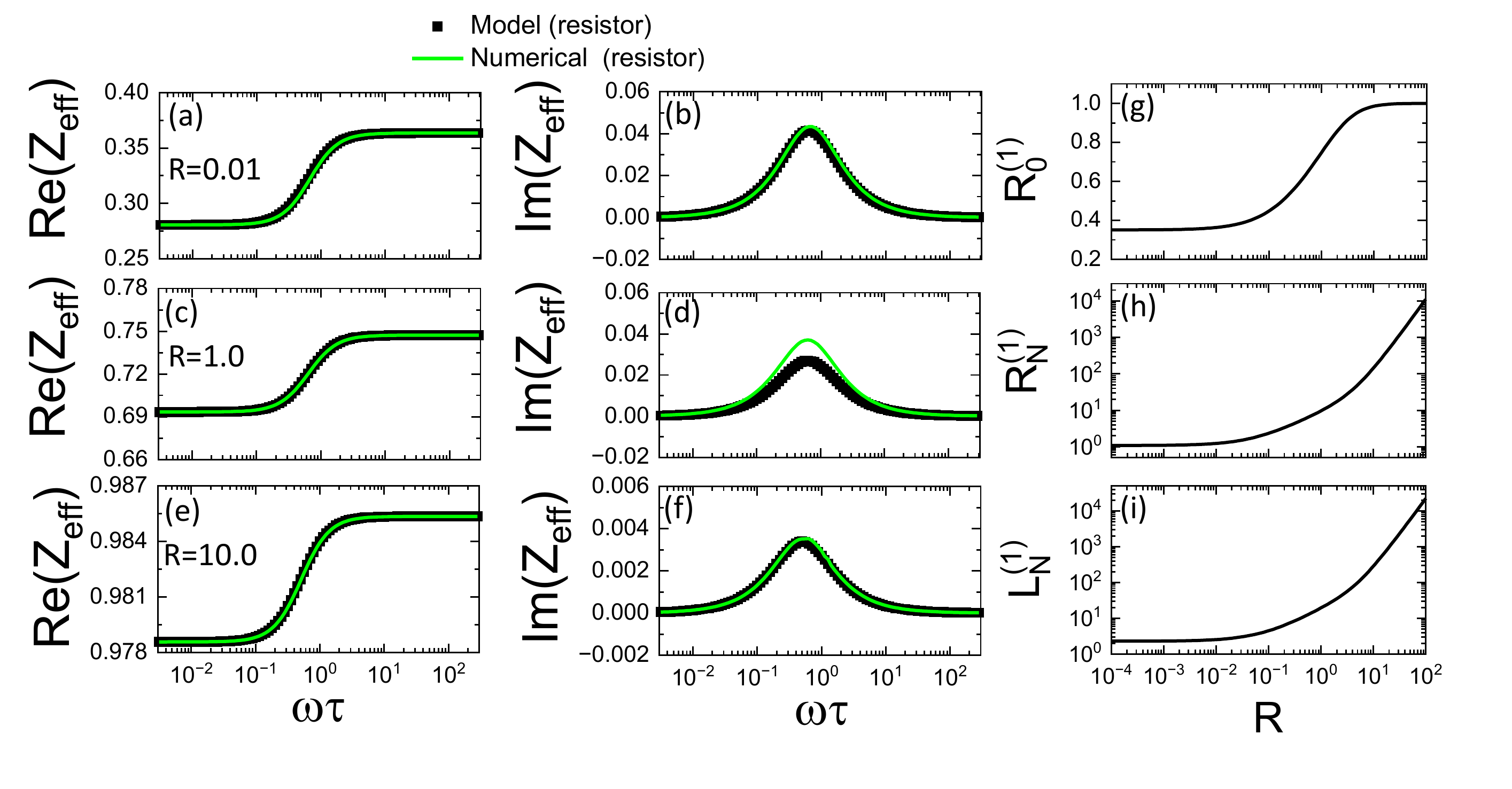} \\
	   \caption{\label{fig:Resistor comparação} Comparison between numerical calculations (green line) and the apparent circuit model (black square) for the memristor associated in series with an external resistor at selected values of resistance. Subfigures (a) and (b) correspond to the real and imaginary parts of the effective impedance, respectively, for $R=0.01$. Subfigures (c) and (d) present the real and imaginary parts of the impedance for $R=1.0$. Subfigures (e) and (f) show the real and imaginary parts of the impedance for $R=10.0$. Subfigures (g), (h), and (i) illustrate the parameters $R_0$, $R_N$ and $L_N$, which are components of the equivalent circuit representation of the memristor response when considering the first Fourier mode.
}
\end{figure}

We first examine the apparent memristor impedance, $Z_{\mathrm{eff}}$, for a memristive device connected in series with a resistor. Figure~\ref{fig:Resistor comparação}(a)-(f) compares the effective impedance extracted from the numerical solution of the complete nonlinear system with that predicted by the apparent equivalent-circuit model. The solid green curves correspond to the numerical evaluation of the effective impedance, obtained from the fundamental harmonic as $Z_{\mathrm{eff}}=Z_{(1)}-R$, for representative resistance values of $R=0.01$, $1.0$, and $10$. The black symbols represent the corresponding predictions of the apparent circuit, calculated from Eqs.~\ref{eq:imp_real} and \ref{eq:imp_im}, after determining the effective parameters through the asymptotic relations given by Eqs.~\ref{eq:seff}, \ref{eq:geff}, and \ref{eq:teff}. This comparison provides a direct assessment of whether the composite system can still be represented by the same equivalent-circuit topology once the resistor-induced renormalization of the memristive response is incorporated into the apparent parameters.

For $R=0.01$, shown in Fig.~\ref{fig:Resistor comparação}(a) and (b), the apparent equivalent circuit reproduces both the real and imaginary components of the effective impedance with remarkable accuracy over the entire frequency range. This agreement indicates that, although impedance additivity breaks down at the microscopic level, the influence of the external resistor can be captured through a renormalization of the apparent memristor parameters, preserving the equivalent-circuit description of the fundamental mode. The resulting impedance closely resembles that of an inductive-like branch shown in Fig.~\ref{fig:diagram}(b), characterized by a step-like increase in the real part of the impedance and a positive peak in the imaginary part. These features reflect the existence of a characteristic memory-relaxation timescale, whose frequency-domain signature is analogous to that of an RL network, despite originating from the internal state dynamics of the memristive system rather than from a physical inductance. 

A different behavior emerges for the intermediate resistance $R=1.0$, shown in Fig.~\ref{fig:Resistor comparação}(c) and (d). While the real part of the impedance remains reasonably well described, the imaginary component exhibits noticeable deviations, particularly around its maximum, where the numerical solution predicts a significantly larger peak than the apparent model. This discrepancy indicates that the interaction between the resistor and the internal memristive dynamics becomes sufficiently strong that the first harmonic can no longer be fully captured by a simple renormalization of the isolated memristor parameters. Instead, the coupling generates additional dynamical effects that are beyond the scope of the proposed equivalent circuit and would translate in a phase shift renormalization of the expected impedance for an isolated device.

Interestingly, good agreement is recovered for the large-resistance regime ($R=10$), as illustrated in Fig.~\ref{fig:Resistor comparação}(e) and (f). In this limit, the voltage drop across the memristive element becomes relatively small, weakening its nonlinear dynamics and driving the composite circuit toward an effectively linear response. Consequently, the apparent equivalent circuit again provides an accurate description of the fundamental harmonic.

The evolution of the extracted apparent circuit parameters is summarized in Fig.~\ref{fig:Resistor comparação}(g)--(i). Figure~\ref{fig:Resistor comparação}(g) shows that the effective parallel resistance, $R^{\mathrm{eff}}_0$, remains close to 0.3 for small values of the external resistance before gradually increasing and saturating near $1.0$ for $R\gtrsim 10$. In contrast, the series resistance $R^{\mathrm{eff}}_S$ and the series inductance $L^{\mathrm{eff}}_S$, displayed in panels (h) and (i), remain close to their intrinsic memristor values for small $R$, but increase by nearly four orders of magnitude as the external resistance reaches $R=100$. This pronounced renormalization reflects the progressive suppression of the intrinsic memristive dynamics by the external resistor

These results demonstrate that the extracted effective parameters provide considerably more than a numerical fit. They quantify how the presence of the passive element modifies the apparent properties of the memristor and establish the parameter range over which an equivalent-circuit description of the fundamental harmonic remains valid. Conversely, the deviations observed for intermediate resistance values ($R\sim 1/G_0$) identify the regime in which the breakdown of impedance additivity becomes most significant and cannot be captured by a simple renormalization of circuit parameters.

\begin{figure}[t]
	\includegraphics[width=0.6\textwidth]{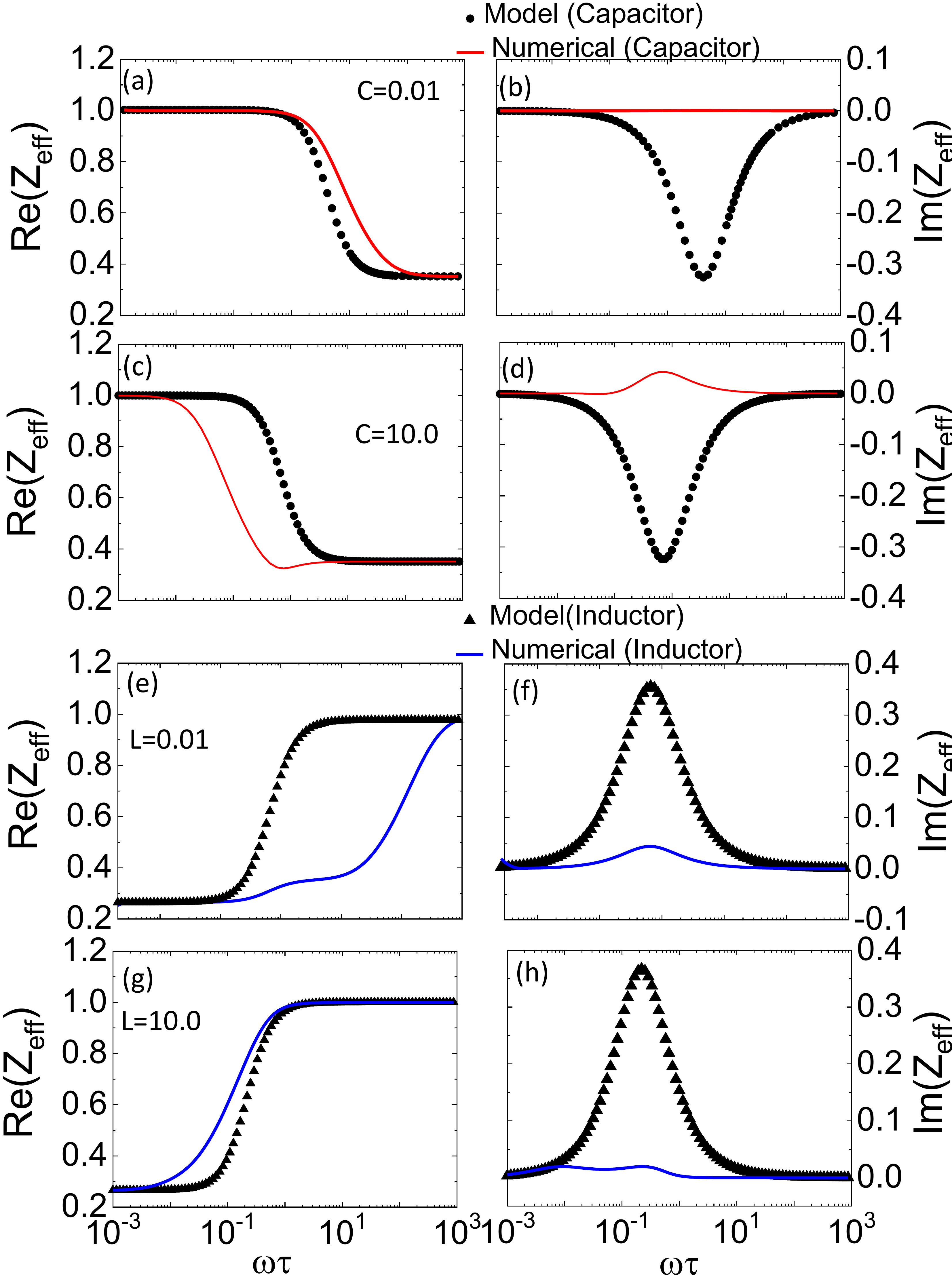} \\
	   \caption{ \label{fig:Capacitor e indutor comparação} Comparison between numerical calculations and the apparent circuit model for the memristor associated in series with external capacitor and inductor. The red lines represent the numerical calculation for the capacitor, while the blue lines correspond to the numerical calculation for the inductor. The black circles denote the theoretical model for the capacitor, and the black triangles denote the theoretical model for the inductor.  Subfigures (a) and (b) correspond to the real and imaginary parts of the effective impedance for $C=0.01$, while subfigures (c) and (d) present the same quantities for $C=10.0$. Subfigures (e) and (f) illustrate the real and imaginary parts of the impedance for $L=0.01$, and subfigures (g) and (h) show the corresponding results for $L=10.0$.
 }
\end{figure}

The same procedure was applied to the memristor connected in series with a capacitor and an inductor, and the resulting effective impedances, $Z_{\mathrm{eff}}=Z_{(1)}-Z_p$, are presented in Fig.~\ref{fig:Capacitor e indutor comparação}. For the memristor--capacitor configuration with a small capacitance ($C=0.01$), shown in Fig.~\ref{fig:Capacitor e indutor comparação}(a) and (b), the apparent equivalent circuit reproduces the real part of the effective impedance reasonably well over the investigated frequency range. The imaginary part, however, exhibits a pronounced negative peak in the apparent circuit calculations that is completely absent from the numerical model. This discrepancy indicates that the coupling between the charge-storage dynamics of the capacitor and the internal state of the memristor generates reactive contributions that cannot be incorporated into a simple renormalization of the isolated memristor parameters. The disagreement becomes even more pronounced for large capacitances ($C=10$), as shown in panels (c) and (d), where neither the real nor the imaginary component of the effective impedance can be satisfactorily reproduced by the apparent circuit. A similar conclusion is reached for the memristor--inductor configuration. For small inductance ($L=0.01$), displayed in Fig.~\ref{fig:Capacitor e indutor comparação}(e) and (f), the effective model fails to reproduce both the real and imaginary parts of the relative impedance. Increasing the inductance to (L=10), Figs. 4(g) and (h), improves the agreement for the real component, whereas substantial discrepancies remain in the imaginary part. Altogether, the results of Figs.~\ref{fig:Resistor comparação} and \ref{fig:Capacitor e indutor comparação} demonstrate that, although an apparent equivalent circuit may adequately reproduce the fundamental response in selected parameter regimes, the impedance of a memristor coupled to passive elements cannot, in general, be obtained from the conventional series-association rule. The interaction between the passive element and the memory dynamics gives rise to a renormalized response that depends on the specific coupling mechanism and cannot be represented by a simple addition of independent impedances.

\begin{figure}[t]
	\includegraphics[width=0.6\textwidth]{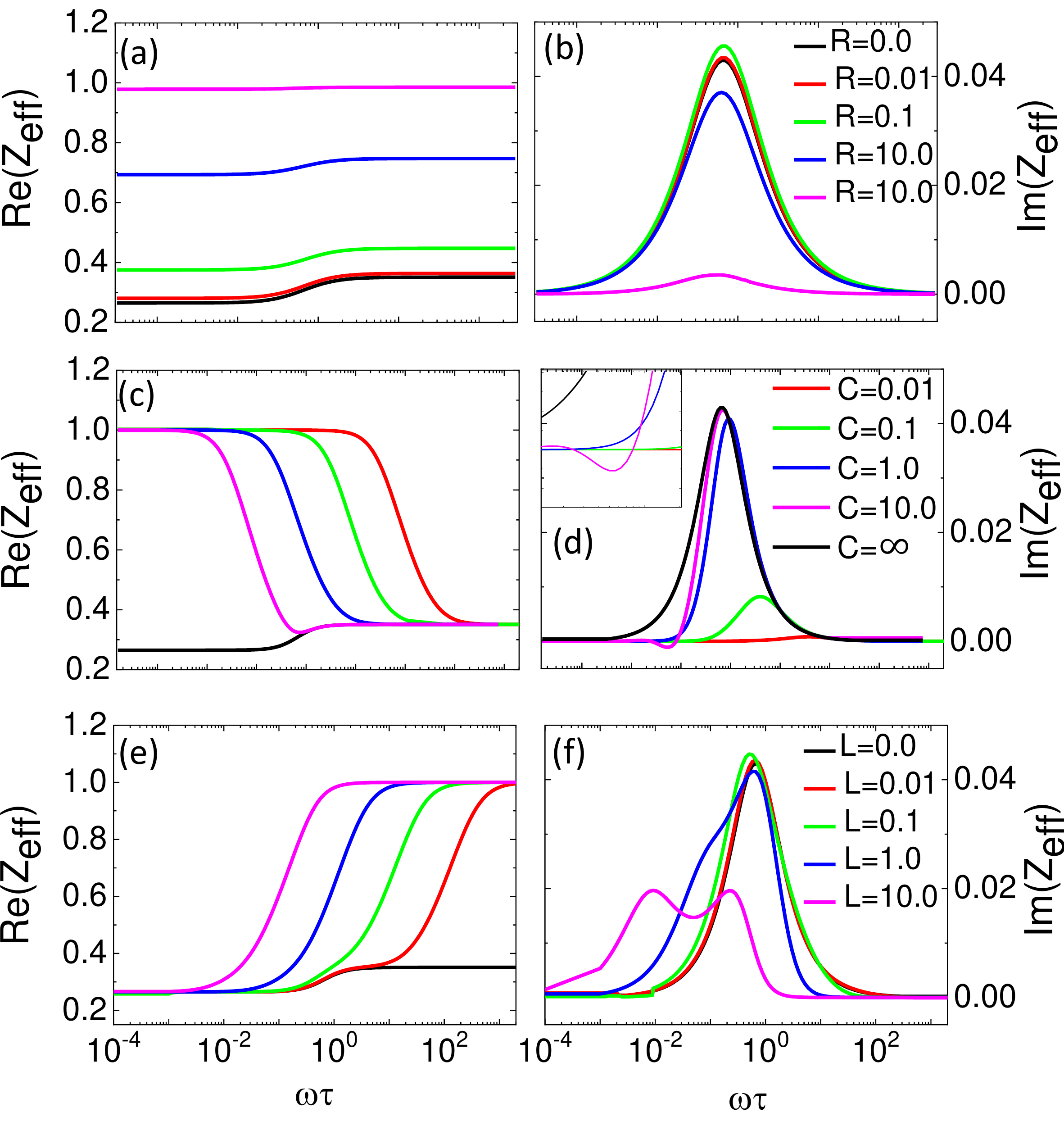} \\
	   \caption{ \label{fig:numerical Bode} Bode-diagram representation of the spectral analysis of the memristor in series with external passive elements (resistor, capacitor, and inductor). In all panels, the black curve represents the isolated memristor (i.e., without any external element), serving as a reference for comparison. Subfigures (a) and (b) correspond to the case of the memristor connected in series with an external resistor, where the parameter $R$ is varied in steps of $10$ from $R = 0.01$ up to $R = 10.0$. Specifically, (a) shows the real part of the effective impedance, while (b) shows the imaginary part. Subfigures (c) and (d) illustrate the case of the memristor connected in series with an external capacitor, with the parameter $C$ varied in steps of $10$ from $C = 0.01$ up to $C = 10.0$; (c) presents the real part of the impedance, while (d) presents the imaginary part. Subfigures (e) and (f) depict the case of the memristor connected in series with an external inductor, where the parameter $L$ is varied in steps of $10$ from $L = 0.01$ up to $L = 10.0$; (e) shows the real part of the impedance, while (f) shows the imaginary part.}
\end{figure}

\begin{figure}[t]
	\includegraphics[width=0.7\textwidth]{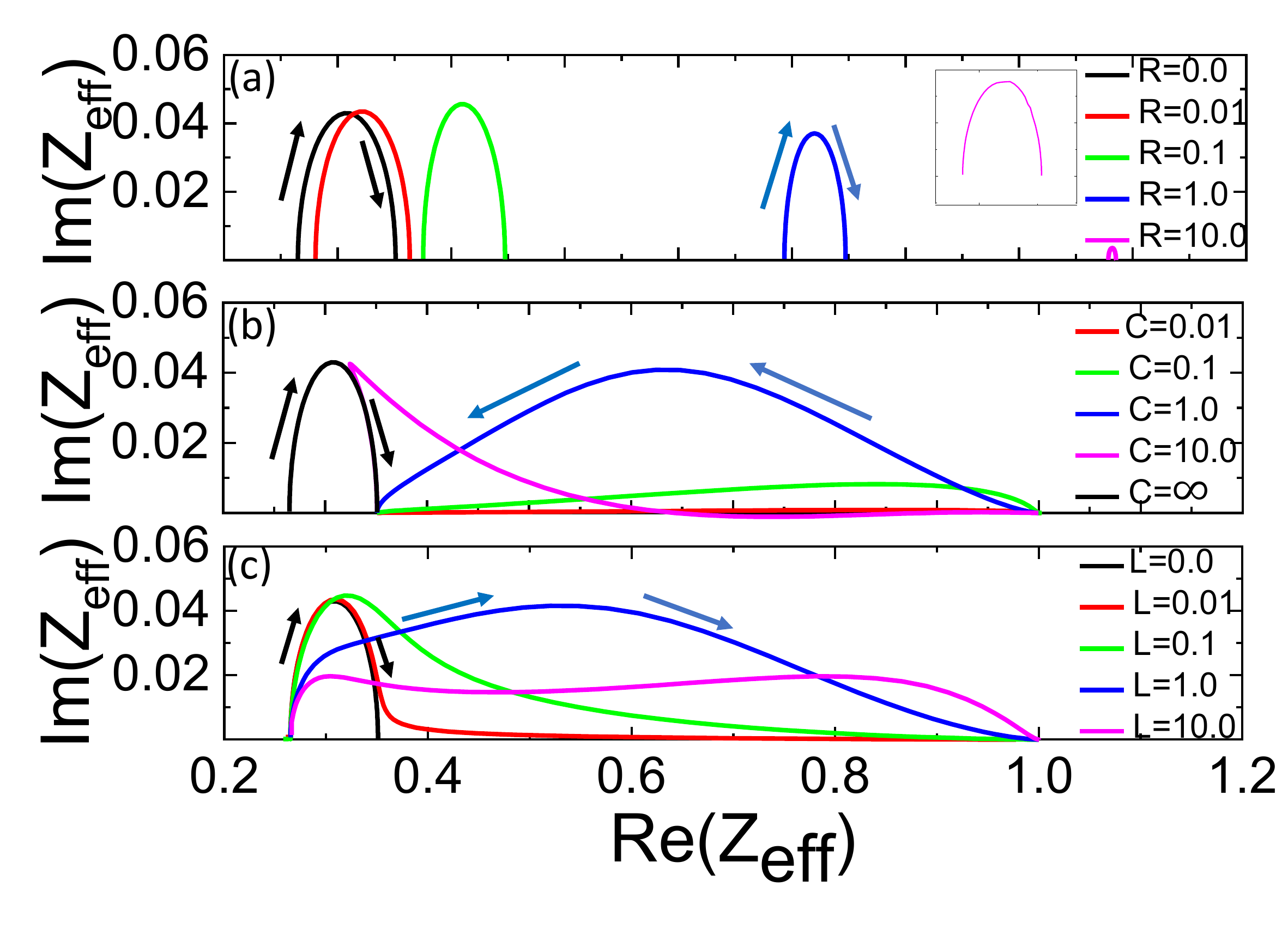} \\
	   \caption{ \label{fig:numerical Nyquist} Nyquist-diagram representation of the spectral analysis of the memristor in series with external passive elements (resistor, capacitor, and inductor). In all panels, the black curve represents the isolated memristor (i.e., without any external element), serving as a reference for comparison, with arrows indicating the direction of increasing frequency. Subfigure (a) correspond to the case of the memristor connected in series with an external resistor, where the parameter $R$ is varied in steps of $10$ from $R = 0.01$ up to $R = 10.0$. Subfigure (b)  illustrate the case of the memristor connected in series with an external capacitor, with the parameter $C$ varied in steps of $10$ from $C = 0.01$ up to $C = 10.0$. Subfigures (c) depict the case of the memristor connected in series with an external inductor, where the parameter $L$ is varied in steps of $10$ from $L = 0.01$ up to $L = 10.0$.  }
\end{figure}

The evolution of the effective impedance with the passive-element parameters is further illustrated through the Bode and Nyquist representations shown in Figs.~\ref{fig:numerical Bode} and \ref{fig:numerical Nyquist}. For the resistor-memristor configuration [Figs.~\ref{fig:numerical Bode} (a) and (b)], increasing the external resistance progressively shifts the real part of the effective impedance to larger values, while the imaginary component exhibits a non-monotonic behavior: it increases for small resistances ($R=0.01$ and 0.1) but decreases once the resistor becomes the dominant element ($R>1$). The corresponding Nyquist diagram, shown in Fig.~\ref{fig:numerical Nyquist}(a), reveals a continuous evolution of the apparent relaxation process as the resistor increasingly suppresses the intrinsic memristive dynamics.

For the memristor-capacitor system [Figs.~\ref{fig:numerical Bode}(c) and (d)], the limiting case $C\rightarrow\infty$ corresponds to the isolated memristor and reproduces the characteristic step-like variation of the real part of the impedance on a logarithmic frequency scale. The midpoint of this transition occurs at $\omega\tau=1/2$, coinciding with the condition for maximum hysteresis area previously identified for type-II memristive responses~\cite{2023-LopezRichard}. Finite capacitances substantially modify this behavior. The real part of the effective impedance approaches the isolated-memristor limit only at sufficiently high frequencies, whereas the low-frequency step becomes inverted, demonstrating that the external capacitor fundamentally alters the relaxation dynamics rather than merely contributing an additive reactance. The imaginary part exhibits an even stronger departure from the isolated-device response. Although convergence is again observed at high frequencies, the vicinity of $\omega\tau\approx1/2$ is characterized by dramatic changes, including a complete sign reversal for $C=10$, as highlighted in the inset of Fig.~\ref{fig:numerical Bode}(d). These modifications are clearly reflected in the Nyquist representation [Fig.~\ref{fig:numerical Nyquist}(b)], where every finite capacitance produces trajectories that differ qualitatively from that of the isolated memristor.

The results for the memristor-inductor configuration are presented in Figs.~\ref{fig:numerical Bode}(e) and (f), together with the corresponding Nyquist diagram in Fig.~\ref{fig:numerical Nyquist}(c). In contrast to the capacitive case, the real part of the effective impedance approaches the isolated-memristor limit ($L=0$) only in the low-frequency regime, with the agreement improving as the inductance decreases. The imaginary component is affected much more strongly by the coupling to the inductor and differs substantially from the isolated-device response for all but the smallest inductance considered ($L=0.01$). Particularly noteworthy is the emergence of two distinct maxima for $L=10$, indicating the appearance of a secondary relaxation process in the effective reactance. This additional feature has no counterpart in the isolated memristor and provides clear evidence that the coupling between the inductive dynamics and the internal memory state generates new characteristic timescales that cannot be captured within the framework of conventional impedance addition.

\section{Conclusion}

We have investigated the extent to which the conventional principle of impedance additivity remains applicable to circuits composed of a memristive device connected in series with passive linear elements. By analyzing resistor-memristor, capacitor-memristor, and inductor-memristor configurations under periodic excitation, we demonstrate that the interaction between the internal memory dynamics and the passive component fundamentally modifies the circuit response. As a consequence, the impedance of the composite system cannot, in general, be obtained by the simple addition of the individual impedances, even when only the fundamental harmonic is considered.

To quantify this breakdown of the superposition principle, we introduced the concept of an apparent equivalent circuit, whose effective parameters are extracted directly from the frequency response of the composite system. Rather than representing intrinsic properties of the memristive device, these apparent parameters characterize how the passive element renormalizes the observed electrical response. Their evolution provides a quantitative measure of the validity of equivalent-circuit descriptions and identifies the parameter regimes in which they remain reliable approximations.

Our results further reveal that the accuracy of equivalent-circuit representations depends strongly on both the nature and the magnitude of the passive element. Resistive, capacitive and inductive couplings generate additional reactive dynamics that cannot be captured by a simple renormalization of the isolated memristor parameters. These deviations manifest through modifications of the Bode and Nyquist characteristics, enhanced higher-order harmonics, and the emergence of additional relaxation features, all of which constitute clear signatures of the failure of impedance additivity.

Beyond memristive systems, the present framework is applicable to any electronic device exhibiting conductive memory, including electrochemical, ionic, dielectric, and photovoltaic systems operating far from equilibrium. We therefore expect these results to contribute not only to the interpretation of impedance spectroscopy in nonlinear systems, but also to the development of more robust equivalent-circuit models and characterization protocols for emerging memory-enabled electronic devices.

\begin{acknowledgments}

This study was financed in part
by the Coordenação de Aperfeiçoamento de Pessoal de
Nível Superior - Brazil (CAPES) and the Conselho Nacional de Desenvolvimento Científico e Tecnológico -
Brazil (CNPq) (Projs. 311536/2022-0, 141841/2023-0) and FAPESP (Projs. 2024/09298-7, 2025/04805-0, 2025/23224-9)

\end{acknowledgments}

\section{Author Declarations}

\subsection{Conflict of Interest}

The authors have no conflicts to disclose.

\subsection{Author Contributions}

Luiz A. Meneghetti Jr.: Conceptualization, Methodology, Software, 
Formal analysis, Investigation, Visualization, and Writing -- original draft.

Leonardo K. Castelano: Conceptualization, Methodology, Supervision, Visualization, Validation, and Writing -- review and editing.

Antônio Sérgio dos Santos: Conceptualization, Investigation, Visualization, Validation, and Writing --review and editing.

Soumen Pradhan: Conceptualization, Investigation, Visualization, Validation, and Writing --review and editing.

Fabian Hartmann: Conceptualization, Investigation, Visualization, Validation, and Writing --review and editing.

Ovidiu Lipan: Conceptualization, Investigation, Visualization, Validation, and Writing --review and editing.

Sven Hofling: Conceptualization, Investigation, Visualization, Validation, and Writing --review and editing. 

Victor Lopez-Richard: Conceptualization, Methodology, Supervision, Visualization, Validation, and Writing -- review and editing.

\section{Data Availability}

The data that support the findings of this study are available from 
the corresponding author upon reasonable request.


%
%

%


\bibliography{bibli}

\end{document}